\documentstyle[preprint,aps,pra,psfig]{revtex}
\begin{document}
\draft
\title{Microscopic Treatment of Binary Interactions in the Non-Equilibrium
Dynamics of Partially Bose-condensed Trapped Gases}
\author{N.P. Proukakis  and K. Burnett}.
\address{Clarendon Laboratory, Department of Physics, University of Oxford,\\
Oxford, OX1 3PU, United Kingdom. }
\author{H.T.C. Stoof}
\address{University of Utrecht, Institute for Theoretical Physics,
Princetonplein 5,\\ P.O. Box 80.006, 3508 TA Utrecht, The Netherlands.}
\date{\today}
\maketitle

\begin{abstract}

In this paper we use microscopic arguments to derive a nonlinear
Schr\"{o}dinger equation for trapped Bose-condensed gases. This is made
possible by considering the equations of motion of various anomalous averages.
The resulting equation explicitly includes the effect of repeated binary
interactions (in particular ladders) between the atoms. Moreover, under the
conditions that dressing of the intermediate states of a collision can be
ignored, this equation is shown to reduce to the conventional Gross-Pitaevskii
equation in the pseudopotential limit. Extending the treatment, we show first
how the occupation of excited (bare particle) states affects the collisions,
and thus obtain the many-body T-matrix approximation in a trap. In addition, we
discuss how the bare particle many-body T-matrix gets dressed by mean fields
due to condensed and excited atoms. We conclude that the most commonly used
version of the Gross-Pitaevskii equation can only be put on a microscopic basis
for a restrictive range of conditions. For partial condensation, we need to
take account of interactions between condensed and excited atoms, which, in a
consistent formulation, should also be expressed in terms of the many-body
T-matrix. This can be achieved by considering fluctuations around the
condensate mean field beyond those included in the conventional finite
temperature mean field, i.e.\ Hartree-Fock-Bogoliubov (HFB),
theory.
\end{abstract}

\section{Introduction}

The observation of Bose-Einstein condensation (BEC) in alkalis \cite{JILA} -
\cite{Rice} strongly motivates a description of the evolution of the condensate
that takes full account of the microscopic nature of atomic interactions in a
trap, both close to and far from equilibrium. The conventional description
relies heavily on the well-known Gross-Pitaevskii Equation (GPE) \cite{NLSE}, also known as the nonlinear Schr\"{o}dinger equation. In this equation, one assumes that the atoms are all effectively condensed and the atomic interactions can be
accurately modeled by a pseudopotential, expressed in terms of the s-wave
scattering length. This theory appears to make good predictions about the
condensate's properties \cite{keith,Pete,Ed,Holland,Baym,NIST_Exc_Comp,Lin_Resp,Fett,Dalfovo,NIST_Vort_1,NIST_Exc_Rev,Dal_NIST,Exc} and it is obviously desirable
to put this phenomenological theory on a clear microscopic basis. In fact, in
spite of its frequent use, a direct link of this effective interaction approach
to microscopic properties of the gas, and a discussion of the effect of the mean
fields on the intermediate states of a binary collision, appears to be lacking.
In this paper we will address such issues in a  derivation of a nonlinear
Schr\"{o}dinger equation based on these microscopic properties and what we
believe are reasonable assumptions about them. In particular, we shall show how
to take account of all possible repeated binary atomic collisions in the
presence of a condensate, thus deriving a mean field evolution in terms of the
many-body Transition (or simply T) matrix. 

Our approach is not limited to cases
close to equilibrium, and therefore complements earlier work in this area.
For example, Bijlsma et al. \cite{Stoof_Var} have
used a variational approach to calculate equilibrium
properties, such as the normal and anomalous self-energies, in the
many-body T-matrix approximation. In addition to this work, one of us has employed a
functional formulation of the Keldysh theory to derive the time-dependent
Landau-Ginzburg theory for the long-wavelength dynamics of an inhomogeneous
weakly-interacting gas at nonzero temperatures, also in the many-body T-matrix approximation
\cite{Nucl}. In this paper we will show, by different methods,
how to derive a time-dependent generalized nonlinear
Schr\"{o}dinger equation\footnote{To avoid confusion, we shall henceforth refer
to the conventionally used equation as the GPE and the one we shall be deriving
here as the NLSE.} in a trap for both zero and nonzero temperatures based on
microscopic arguments. We shall explain that, to obtain this equation, we must neglect the effect of mean fields in the intermediate collisional states; we shall also discuss the physical significance of these effects.

In an earlier publication \cite{Prou1}, we derived equations of motion for
thermal averages of products of up to three (single-particle) fluctuation
operators. In this way, we obtained a time-dependent version \cite{Foot2,Griffin_HFB} of the Hartree-Fock-Bogoliubov
(HFB) equations \cite{HFB1,Dorre,Kobe,Blaizot} in terms of actual interatomic
potentials, and further generalized them by considering more complex anomalous
averages (triplets).
This closed system of equations can be used in two different ways: In the first approach, all averages of
products of fluctuation operators evolve on similar timescales, so that the
equations need to be solved self-consistently. Such a treatment enables us to
investigate the possibility of further order parameters being present in our
system. We point out that the possibility of pairing as a competing transition to BEC in
the case of attractive interactions has already been investigated in
\cite{Stoof_Attr}. The equations of \cite{Prou1} further allow for the
possibility of three atoms grouping together, such as for example, condensation
of triplets \cite{Laloe}. However, there are also situations (in particular a
dilute gas with repulsive interactions), in which the higher order (anomalous)
correlations vary on  faster timescales, enabling us to formally eliminate them
from the equations of motion, i.e. by integrating over their effect during
collisions. In this paper we will show how this formal (adiabatic) elimination
of the pair correlation in our HFB equations gives rise to a nonlinear
Schr\"{o}dinger equation that includes the non-dressed repeated binary
interactions (ladders) at zero temperature. We shall furthermore extend our
treatment to nonzero temperatures, for which we shall also need to eliminate
the triplets mentioned above.

The equations of motion for averages of products of (up to three) single-particle fluctuation operators have been given in appendix A, although we refer the reader to  \cite{Prou1} for
more details. We emphasize that these equations do not merely bring the bare
particle ladder interactions into our formalism, but also include more complex
effects, such as  dressing and damping of intermediate states during collisions.
In this paper, we discuss how --- and in what limits --- these microscopic
equations reduce to the conventionally used phenomenological Gross-Pitaevskii
expression. We also discuss the possibility of consistent theories outside
these limits. Furthermore, we explicitly mention what processes must be
neglected in order to obtain a NLSE. In fact, we shall see one has to neglect
the effect of the condensate mean field and those due to the presence of
excited atoms on the intermediate states of a binary collision \cite{Prou_RoyalSoc}. Identification
of these terms shows how to explore the deviation from the bare-particle
T-matrix due to dressing generated by the mean fields. This treatment enables
us to make qualitative predictions about the validity regime of the
phenomenological GPE, an issue we hope to address computationally for
inhomogeneous gases in the future.

One might expect the zero-temperature evolution of the mean field of a trapped
Bose-Einstein condensate  in the
simplest version of the mean field theory, to be described --- in the occupation number representation --- by 
\begin{equation}
i \hbar \frac{dz_{n}(t)}{dt}  = \sum_{k}h_{nk}^{(0)}(t)z_{k}(t)~.
\end{equation}
Here $z_{n}(t)$ corresponds to the time-dependent mean value amplitude of the
$n^{th}$ trap level which is obtained from the single-particle operators
$\hat{a}_{n}(t)$ according to the shift \cite{Blaizot}
\begin{equation}
\hat{a}_{n}(t) = z_{n}(t) + \hat{c}_{n}(t)~.
\end{equation}
Here the operators $\hat{a}_{n}$ are defined by the usual decomposition of the
Bose field operator $\hat{\Psi}({\bf r},t)$ into any complete set of
orthonormal single-particle states  $\psi_{n}({\bf r})$, namely
\begin{equation}
\hat{\Psi}({\bf r},t) = \sum_{n} \psi_{n}({\bf r})
                                 \hat{a}_{n}(t)~.
\end{equation}
In this simplest approach, the time-dependent condensate mean field depends on the Hartree-Fock hamiltonian ${\bf h^{(0)}}(t)$ given by
\begin{equation}
h_{nk}^{(0)}(t) =  \langle n | \hat{\Xi} | k \rangle   + \sum_{ij} \langle ni|
\hat{V} | jk \rangle z_{i}^{*}(t)z_{j}(t)~.
\end{equation}
Here $\hat{\Xi}$ contains the kinetic energy and trap potential and $\langle
ni| \hat{V} | jk \rangle$ represents the symmetrized form of the actual
(single-vertex) interatomic potential \cite{Foot3} between a pair of colliding particles. This is  defined in terms of
$\psi_{n}({\bf r})$  by
\begin{equation}
\langle ni| \hat{V} |  jk \rangle = \frac{1}{2} \left\{ \left( ni| \hat{V} |
jk \right) + \left( ni| \hat{V} |  kj \right) \right\}~,
\end{equation}
where
\begin{equation}
\left( ni| \hat{V} |  jk \right) = \int \int d{\bf r} d{\bf r^{'}}
 \psi_{n}^{*}({\bf r})  \psi_{i}^{*}({\bf r^{'}})  V ({\bf r} - {\bf r^{'}})
\psi_{k}({\bf r^{'}})  \psi_{j}({\bf r}) \end{equation}
and $V({\bf r} - {\bf r^{'}})$ represents the actual interatomic potential
experienced between two interacting atoms at {\em each} collisional vertex. ¦
Although Eq. (1) may on first sight appear to be equivalent to the
conventional form of the GPE \cite{NLSE}
\begin{equation}
i \hbar \frac{\partial \Phi({\bf r} ,t)}{\partial t} = \left( - \frac{\hbar^{2}
\nabla_{{\bf r}}^{2}}{2m} + V_{trap}({\bf r}) \right) \Phi({\bf r}, t ) +
NU_{0}  | \Phi({\bf r} ,t) |^{2}  \Phi({\bf r} ,t)~,
\end{equation}
this is not the case. The GPE takes account of all (repeated) collisional
processes via a pseudopotential of the form $V({\bf r} - {\bf r} ^{'}) =  U_{0}
\delta({\bf r} - {\bf r} ^{'})$ where $U_{0} = 4 \pi \hbar^{2} a/m$
\cite{Fermi}. On the contrary, the matrix element defined in (5) represents the
{\em instantaneous} (i.e single-vertex) interaction between two atoms as
depicted in Fig. 1(a).

In this paper we will show that we can include the ladder diagrams of Fig.
1(b) into our formalism, by considering the evolution of correlations of
products of fluctuation operators $\hat{c}_{n}$ which are given in appendix A.
These correlations influence the condensate mean field according to the
exact equation \cite{Prou1}
\begin{equation}
i \hbar \frac{dz_{n}}{dt}  =  \sum_{k} \langle n | \hat{\Xi} | k \rangle z_{k}
+ \sum_{ijk} \langle ni| \hat{V} | jk \rangle \left[ z_{i}^{*}z_{j}z_{k} +
\kappa_{jk}z_{i}^{*} + 2 \rho_{ji}z_{k} + \lambda_{ijk} \right]~,
\end{equation}
where we have defined the quantities
\begin{equation}
\rho_{ji} = \langle \hat{c}_{i}^{\dag} \hat{c}_{j} \rangle~, \hspace{1.5cm}
\kappa_{jk} = \langle \hat{c}_{k} \hat{c}_{j} \rangle = \langle \hat{c}_{j} \hat{c}_{k} \rangle~, \hspace{1.5cm}
\lambda_{ijk} = \langle \hat{c}_{i}^{\dag} \hat{c}_{j} \hat{c}_{k} \rangle =
\langle \hat{c}_{k} \hat{c}_{j} \hat{c}_{i}^{\dag}  \rangle~.
\end{equation}
In obtaining (8) we have merely assumed the existence of a mean field (in the
sense of Eq. (2)) and that the dominant collisions in our atomic assembly
occur pairwise.

The main part of this paper deals with the microscopic derivation of a
nonlinear Schr\"{o}dinger equation for bare particles in a trap, and we shall
discuss in what limits it reduces to simpler versions. The use of a NLSE
necessitates a well-defined and thus slowly-varying condensate fraction, which
in our formalism corresponds to a slowly-evolving mean field amplitude $z_{n}$,
in comparison with the duration of a typical binary interaction. This means
that, in most of the analysis given below, we are limiting ourselves to the
case of net repulsive interactions between the atoms, i.e. we are not allowing
for BCS-type effects \cite{Stoof_Attr} to become important.
When extending our treatment to nonzero temperatures, we will similarly need to
assume slow evolution of the excited state populations, although this issue
becomes rather subtle as we shall discuss in Sec. III C. In our subsequent
calculations we shall discuss when the thermal correlations $\kappa_{jk}$ and
$\lambda_{ijk}$ vary on a much smaller timescale and show how they can be
formally eliminated in our treatment.

This paper is divided into five parts. In the first part of the paper we
restrict ourselves to near zero temperatures, for which it appears reasonable to neglect all effects of excited states. We thus obtain a NLSE that includes the full (bare particle) ladder interactions, where this effective interaction potential arises from the adiabatic elimination of the anomalous correlation
$\kappa_{jk}$. This equation is then shown to reduce to the phenomenological
Gross-Pitaevskii equation by means of the usual pseudopotential approximation.
In Sec. III, we extend our treatment to nonzero temperatures. The first
modification that comes about is the inclusion of the occupation of virtually
excited states into the ladder diagrams. However, the finite number of excited
atoms further necessitates the consideration of collisions between condensed
and excited atoms. We thus show how adiabatic elimination of $\lambda_{ijk}$
`upgrades' the condensate-excited state interaction potential to  the ladder
approximation. In Sec. IV, we show how the particle propagators get dressed
by the condensate mean field and by interactions with other excited atoms and
discuss the consequences of identifying these terms in our equations. We then
discuss, in Sec. V, the validity of several mean field theories for both
zero and nonzero temperatures and in particular point out the inconsistency of
the conventional finite temperature mean field approach (i.e.\ what is commonly
referred to as HFB). Finally, in Sec. VI, we discuss some further effects of
the mean fields on collisions. The detailed form of the equations needed for
the current discussion have been reproduced in appendix A, whereas appendix B
gives an exact treatment of the condensate mean field evolution up to second
order in the interaction potential.

\section{Ladder Interactions in a Bare Particle Basis at zero Temperatures}

Let us initially assume that the atomic assembly is sufficiently dilute, so
that we can treat the bare particles as weakly-interacting quantities. In this
limit, bare particles form a reasonable basis set for the system and can thus
be used for the description of the assembly. We shall take the non-interacting
part $\hat{\Xi}$ of the single bare-particle hamiltonian in a trap as diagonal,
by writing
\begin{equation}
\langle n | \hat{\Xi} | k \rangle = \hbar \omega_{n} \delta_{kn}~.
\end{equation}
If this basis accurately represents the state of the system (i.e. all dressing
of states by the mean fields can be ignored), then the anomalous average
$\kappa$ will evolve rapidly, thus enabling us to adiabatically eliminate it.
The elements of $\rho$, however, vary on a slower timescale, as their evolution
depends on an energy difference. This means that they should, in principle, be
retained in the equation of motion (8) for the condensate evolution. Near $T =
0$, however, the occupation of excited states is extremely scarce, so that we
can, in the first instance, neglect its effect on the condensate evolution.

Within these approximations, Eq. (8) reduces to
\begin{equation}
i \hbar \frac{dz_{n}}{dt}  =  \hbar \omega_{n} z_{n} + \sum_{ijk} \langle ni|
\hat{V} | jk \rangle \left[ z_{i}^{*}z_{j}z_{k} + \kappa_{jk}z_{i}^{*}
\right]~.
\end{equation}

In order to adiabatically eliminate the pairing $\kappa_{jk}$, we shall need
its equation of motion given  in appendix A. We note that in a bare particle
basis at $T = 0$, Eq. (A6) essentially simplifies to
\begin{equation}
i \hbar \frac{d \kappa_{kj}}{dt}  = \hbar (\omega_{k} + \omega_{j})\kappa_{kj}
+ \Delta_{kj}~,
\end{equation}
where we have defined \cite{Blaizot}
\begin{equation}
\Delta_{kj} = \sum_{ms} \langle kj| \hat{V} | ms \rangle \left[ z_{m} z_{s} +
\kappa_{ms}  \right]~.
\end{equation}

We are now in a position to derive a NLSE valid at $T = 0$ with the actual
interatomic potential of Eq. (4) replaced by the two-body T-matrix. Such
an equation will be shown to be equivalent to the conventional GPE if we
replace the T-matrix by the zero-range pseudopotential.

Integrating (12) and substituting into (11), we obtain
\begin{eqnarray}
i \hbar \frac{dz_{n}}{dt} & = & \hbar \omega_{n} z_{n} + \sum_{ims} \langle ni|
\hat{V} | ms \rangle  z_{i}^{*}z_{m}z_{s} + \sum_{ijk} \langle ni| \hat{V} | jk
\rangle  z_{i}^{*}(t) \int \frac{dt^{'}}{i} e^{-i(\omega_{k} + \omega_{j})(t -
t^{'})}  \nonumber \\
&  & \times \sum_{ms} \langle jk | \hat{V} | ms \rangle \left[
z_{m}(t^{'})z_{s}(t^{'}) + \kappa_{ms}(t^{'}) \right]~,
\end{eqnarray}
where the integration is to be carried out in the interval $[t_{0},t]$, in
which the collision takes place. Since the gas is considered sufficiently
dilute, the mean collisional duration is negligible compared to the time
between successive collisions, enabling us to take the limit $t_{0} \rightarrow
- \infty$. In writing Eq. (14) we have made use of the approximation that the
pairing $\kappa$ asymptotically relaxes to zero, so that $\kappa_{jk}(t_{0})
\rightarrow 0$ as $t_{0} \rightarrow - \infty$. Physically, this is the same as
saying that these anomalous interparticle correlations are dominantly created
within a collision between condensate particles.

A further crucial point to note is that, since we are trying to derive a NLSE,
we are implicitly assuming that the mean value amplitude $z_{i}(t)$ is a slowly
varying quantity, that is, it remains practically constant over the time scale
in which the correlation $\kappa_{jk}$ varies. Thus, we are justified in
allowing the quantities $z_{i}(t^{'})$ present in the integrand of Eq.
(14) to evolve freely with time by setting
\begin{equation}
z_{i}(t^{'}) = z_{i}(t) e^{+i \omega_{i} (t - t^{'})}~.
\end{equation}

If we now intitially limit ourselves to second order in the interaction
potential $V$, we note that we can disregard the $\kappa_{ms}(t^{'})$ term of
(14), as its first contribution will be to order $V^{3}$. Then, Eq. (14)
reduces to
\begin{equation}
i \hbar \frac{dz_{n}}{dt}  =  \hbar \omega_{n} z_{n} + \sum_{ims} \langle ni|
\hat{T}^{2B}(E) | ms \rangle z_{i}^{*}z_{m}z_{s}~,
\end{equation}
which gives us the bare-particle two-body T-matrix in a trap to second order in
the interaction potential $\hat{V}$, namely
\begin{equation}
\hat{T}^{2B}(E)   = \hat{V} + \sum_{jk} \hat{V} | jk \rangle \int \frac{dt^{'}}{i} e^{-i(\omega_{k} + \omega_{j} - E) (t - t^{'})} \langle jk | \hat{V}~.
\end{equation}
Here $E$ represents the energy of the colliding (bare) atoms prior to the
collision in the centre-of-mass frame (i.e. in Eq. (16), $E = \hbar(\omega_{m} +
\omega_{s})$), whereas $j$ and $k$ correspond to intermediate trap eigenstates
induced during the collision.
However, we point out that it is precisely the presence of the term
$\kappa_{ms}(t^{'})$ which generates $T^{2B}$ to all orders in $V$, as can
easily be seen by repeated use of equations (11)-(13). We have hence shown
Eq. (16) to be valid to all orders in $V$, with $\hat{T}^{2B}$ satisfying
the following Lippmann-Schwinger relation in a trap
\begin{equation}
\hat{T}^{2B} (E) = \hat{V}  + \sum_{jk} \hat{V} | jk \rangle \int \frac{dt^{'}}{i} e^{-i(\omega_{k} + \omega_{j} -  E) (t - t^{'})} \langle jk |
\hat{T}^{2B}(E)~.
\end{equation}
This modified form of the NLSE accurately represents a collision between two
condensate atoms occuring in vacuum, by taking account of all possible states
via which the collision can proceed. Thus, (18) includes the ladder diagrams
for bare particles as depicted diagrammatically in Fig. 2(a).

At this point, we could introduce the many-body T-matrix, by making use of the
next set of terms in $d \kappa/d t$ (given in Eq. (A6)). However, before returning
to this in Sec. III (A), let us focus on the relation between this version
of the microscopically derived zero-temperature NLSE (16) and the
phenomenological GPE (7).

\subsection{Reduction to the Gross-Pitaevskii Equation}

Let us now show explicitly how we can obtain the GPE from our NLSE (16). To do
this we shall need to approximate the (bare-particle) T-matrix defined in (18)
in terms of actual interatomic potentials, by a phenomenological model, such as
the pseudopotential approach discussed in detail in the classic paper by Huang
and Yang \cite{Huang_Yang}. For the low condensation temperatures involved, we
can limit ourselves to the lowest order contribution of the s-wave. We may thus
write
 \begin{equation}
\hat{T}^{2B}({\bf r}) \simeq \frac{4 \pi \hbar^{2} a}{m} \delta({\bf r})~,
\end{equation}
where $a$ corresponds to the s-wave scattering length.

Let us now use this approximation in (16) to explicitly reconstruct Eq. (7). In
doing this, we shall also multiply the resulting equation by $\phi_{n}({\bf
r})/\sqrt{N}$ and sum over the indices $n$ (so as to obtain the desired
$\Phi({\bf r},t)$ on the LHS of our equation). We note that the factor of
$\sqrt{N}$ is needed so as to renormalize the wavefunctions, as the $z_{i}$
satisfy
\begin{equation}
\sum_{i} z_{i}^{*} z_{i} = N~.
\end{equation}
After carrying out the integration with respect to the $\delta$-function and
rearranging, we thus obtain
\begin{eqnarray}
i \hbar \frac{\partial}{\partial t} \Phi({\bf r},t) & & = \int  d{\bf r^{'}}
\left[ \sum_{n} \phi_{n}^{*}({\bf r^{'}}) \phi_{n}({\bf r}) \right]
\hat{\Xi}({\bf r^{'}})   \sum_{k} \phi_{k}({\bf r^{'}}) \frac{z_{k}(t)}{\sqrt{N}}
\nonumber \\
& + & N U_{0} \int d{\bf r^{'}} \left[ \sum_{n}  \phi_{n}^{*}({\bf r^{'}})
\phi_{n}({\bf r}) \right] \sum_{i} \phi_{i}^{*}({\bf r^{'}})
\frac{z_{i}^{*}(t)}{\sqrt{N}}  \sum_{j} \phi_{j}({\bf r^{'}})
\frac{z_{j}(t)}{\sqrt{N}} \sum_{k} \phi_{k}({\bf r^{'}})
\frac{z_{k}(t)}{\sqrt{N}}~.
\end{eqnarray}
Using the completeness relation for orthogonal states, this reduces to the GPE
(7) for $\Phi({\bf r},t)$.

We shall now try to discuss the potential limitations of the GPE (7). To do
this, let us briefly outline its conventional derivation, in which one starts
with the Heisenberg equation of motion for the Bose field operator $\hat{\Psi}
({\bf r}, t)$ in the limit of pairwise interactions. After using the
appropriate commutation relations for the field operators and taking mean
values, we are left with
\begin{eqnarray}
i \hbar \frac{\partial \Phi ({\bf r},t)} {\partial t} & = &  \left( -
\frac{\hbar^{2} \nabla_{{\bf r}}^{2}}{2m} + V_{trap}({\bf r}) \right)
\Phi({\bf r},t) \\ \nonumber
&  & + \int d {\bf r^{'}} V({\bf r} - {\bf r^{'}}) \langle \hat{\Psi}^{\dag}({\bf
r^{'}},t) \hat{\Psi}({\bf r^{'}},t) \hat{\Psi}({\bf r},t) \rangle~,
\end{eqnarray}
where $V({\bf r} - {\bf r^{'}})$ again corresponds to the actual potential
experienced by a pair of interacting atoms (assuming a central potential).

It is at this stage that the standard approximations are conventionally made
(see e.g. \cite{Nozieres}). One first uses the pseudopotential approximation
$V({\bf r} - {\bf r^{'}}) = U_{0} \delta ({\bf r} - {\bf r^{'}})$ and drops the
interatomic correlations by setting $\langle \hat{\Psi}^{\dag}({\bf r^{'}},t)
\hat{\Psi}({\bf r^{'}},t) \hat{\Psi}({\bf r},t) \rangle = \langle
\hat{\Psi}^{\dag}({\bf r^{'}},t) \rangle \langle \hat{\Psi}({\bf r^{'}},t) \rangle
\langle \hat{\Psi}({\bf r},t) \rangle$. The two steps are interlinked and
one could not carry them out independently as the resulting expressions would
not be of any use. The second approximation is usually justified by arguing
that the effect of all correlations thus neglected is taken into account in the
effective interaction strength $U_{0}$.

However, we stress that an effective interaction treatment can only be
rigorously justified if the pseudopotential approximation is carried out after
the two-body interaction potential has been `upgraded' to the T-matrix, in the
manner shown above. Although the resulting equation in this limit is identical
to the GPE, we  have given it a microscopic basis. Actually, we will use this
basis to show that the terms ignored (in the GPE) are not always negligible.

It is well known in the case of the homogeneous gas, that the use of a
$\delta$-function potential in a self-consistent treatment can lead to
unphysical results, such as no depletion of the condensate \cite{Huang_NIST}.
This is due to the fact that such an approximation already contains an implicit
assumption about the actual interaction operator, namely that it is acting on
free particle states \cite{Huang_Book}. Put differently, the $\delta$-function
potential can only be used in combination with an ultra-violet cut-off in the
theory. One must therefore be cautious of some predictions based on a
zero-range effective interaction potential, unless this has been introduced in
the rigorous manner on the two-body T-matrix (i.e. by means of (19)).

Furthermore, it is worth pointing out that the pseudopotential approximation
(19) explicitly ignores all momentum dependence, which appears in the problem
as higher order terms in the s-wave scattering length $a$. This is an excellent
approximation, valid, in the case of the homogeneous gas if $a/\Lambda \ll 1$,
where $\Lambda$ represents the atomic de Broglie wavelength. However, we stress
that our  treatment is not limited to zero-range potentials and can handle
spatially-dependent interaction effects that are not present in the
phenomenological Gross-Pitaevskii model \cite{You}.

Thus far, in this section, we have only discussed the limitations of the GPE in
terms of the treatment of the spatial and momentum dependence of the atomic
interactions. Unfortunately, this is not the only problem we have to address.
The main issue is that we have derived (16) based on a bare particle basis.
This implies, for example, that we have ignored all dressing effects on the
intermediate collisional states due to the condensate mean field. This
obviously means that such effects are also implicitly ignored in the GPE, and
implies that the GPE is strictly only valid if the atoms are colliding in
vacuum. The reason why these mean field effects cannot be included in a
phenomenological GPE based on the s-wave scattering length $a$, is that, in the
experiments, a is spectroscopically determined in the absence of mean fields
(i.e.\ effectively in vacuum). Our microscopic approach fully treats collisions
in the presence of mean fields, and we believe it provides a natural formalism
for obtaining expressions for both dressing and damping of the GPE
\cite{Prou5}.

\section{Ladder Diagrams at Finite Temperatures in a Bare Particle Basis}

Following the analysis of the previous section, it should be clear to the reader, that the GPE
represents only a limiting case of the equations given in  appendix A, since it
neglects the effect of the medium in which the collisions occur (which becomes
increasingly important as the atomic assembly becomes denser). Before we deal
with these issues, we would, however, first like to extend Eq. (16) to
finite temperatures, which corresponds to a well-posed problem within our
formalism.

When extending our treatment to nonzero temperatures, a variety of new features
arises: firstly, there will now be a non-negligible occupation of excited
states, which leads to a modification in the scattering between two condensed
atoms, as discussed in Sec. A below. At the same time, however, the
evolution of the condensate mean field will also be affected by collisions
between condensed and excited atoms, as will be dealt with in Sec. B.

In this section we discuss the limit of weak interactions, in which the
collisions can be described in terms of bare (i.e. non-dressed) atoms. The
deviation from such a simplistic picture (due to dressing induced by mean
fields) will be dealt with in Sec. IV.

\subsection{The Many-Body T-Matrix for Weakly-interacting Particles}

We shall now deal with the first of the finite temperature effects, namely how
Bose statistics affect the T-matrix, by taking account of the occupation of
low-lying excitations during a collision between two condensate particles. This
effect is expected to be so small at low temperatures, that we have chosen to
neglect it altogether near $T = 0$. However, in order to take account of the
occupation of virtual states accessed during a collision between two condensate
particles, we need to modify Eq. (12) to
\begin{equation}
i \hbar \frac{d \kappa_{kj}}{dt}  = \hbar (\omega_{k} + \omega_{j})\kappa_{kj}
+ \Delta_{kj} + \sum_{r} \left[ \rho_{kr} \Delta_{rj}  + \Delta_{kr}
\rho_{rj}^{*}  \right]~.
\end{equation}

We again formally integrate the above equation and adiabatically eliminate
$\kappa_{jk}$ from Eq. (11). Doing so we obtain, to second order in the interaction
potential, the following equation of motion for the condensate
\begin{eqnarray}
i \hbar \frac{dz_{n}}{dt} & = & \hbar \omega_{n} z_{n} + \sum_{ims} \langle ni|
\hat{V} | ms \rangle  z_{i}^{*}z_{m}z_{s} \nonumber \\
& &  + \sum_{ijk} \langle ni| \hat{V} | jk \rangle  z_{i}^{*}(t)  \int \frac{dt^{'}}{i} e^{-i(\omega_{k} + \omega_{j})(t - t^{'})}
\sum_{ms} \langle jk | \hat{V} | ms \rangle z_{m}(t^{'})z_{s}(t^{'}) \nonumber
\\
& &  + \sum_{ijk} \langle ni| \hat{V} | jk \rangle  z_{i}^{*}(t) \bigg\{
\int \frac{dt^{'}}{i} \sum_{rms} e^{-i(\omega_{k} + \omega_{j})(t - t^{'})}
\nonumber \\
& & \times \langle rk | \hat{V} |ms \rangle  z_{m}(t^{'})z_{s}(t^{'})
\rho_{jr}(t^{'}) + \{ k \leftrightarrow j \} \bigg\}~.
\end{eqnarray}
Here, the notation $ + \{ k \leftrightarrow j \}$ indicates the presence of an
identical term upon interchanging the (intermediate propagation) labels $j$ and
$k$, and thus indicates the presence of both a direct and an exchange term (in
much the same manner as those are included in the symmetrization of the
interaction vertex (5)). Once again the $\kappa$ correlations ensure that this
equation generalizes to all $V$. Hence, $T^{2B}$ in Eq. (16) becomes
replaced by the operator $\hat{t}$, defined by
\begin{eqnarray}
\hat{t}(E) & = & \hat{V}  + \sum_{jk} \hat{V} | jk \rangle \int \frac{dt^{'}}{i}
e^{-i(\omega_{k} + \omega_{j} -  E) (t - t^{'})}  \langle jk | \hat{t}(E)
\nonumber \\
& + & \sum_{rjk} \hat{V} | jk \rangle  \Bigg\{   \int \frac{dt^{'}}{i}
e^{-i(\omega_{k} + \omega_{j} -  E) (t - t^{'})}  \rho_{jr}(t^{'}) \langle rk |
 +  (k \leftrightarrow j)  \Bigg\}   \hat{t}(E)~.
\end{eqnarray}

In the limit of weak interactions, we can treat the elements of $\rho$ as
diagonal, thus corresponding to population of excited states. The off-diagonal
elements will correspond to dressing of the bare atoms and will be discussed in
Sec. IV. In this manner, we arrive at the Lippmann-Schwinger relation for
the many-body T-matrix in a trap in terms of bare particles, in the form
\begin{eqnarray}
\hat{T}^{MB}(E)  = \hat{V} +  \sum_{jk} \hat{V} | jk \rangle \int \frac{dt^{'}}{i}
e^{-i(\omega_{k} + \omega_{j} -  E )(t - t^{'})} & & [ 1 + \rho_{jj}^{0}(t^{'})
+ \rho_{kk}^{0}(t^{'}) ] \nonumber \\ & & \times \langle jk | \hat{T}^{MB}(E)~.
\end{eqnarray}
This definition has also been depicted diagramatically in Figs. 2(b)-(c). The
operator $\hat{T}^{MB}$ of Eq. (26) is linked to $\hat{T}^{2B}$ of Eq. (18) by
the equivalent definition
\begin{eqnarray}
\hat{T}^{MB}(E)  = \hat{T}^{2B}(E) +  \sum_{jk} \hat{T}^{2B}(E)
| jk \rangle \int \frac{dt^{'}}{i} e^{-i(\omega_{k} + \omega_{j} -
  E )(t - t^{'})} & & [ \rho_{jj}^{0}(t^{'}) + \rho_{kk}^{0}(t^{'}) ] \nonumber
\\ & & \times \langle jk | \hat{T}^{MB}(E)~.
\end{eqnarray}

It is easy to see that, in the homogeneous limit, the expression (26) reduces
to the well-known integral relation for the many-body T-matrix
\cite{Stoof_NIST}, namely
\begin{eqnarray}
T^{MB} (k^{'},k,K;E) & = & V(k^{'} - k) + \int \frac{dk^{''}}{(2 \pi)^{3}}
V(k^{'} - k^{''}) \nonumber \\
& \times & \frac{ 1 + N(\frac{K}{2} + k^{''}) + N(\frac{K}{2} - k^{''}) }
{E - \frac{ (\hbar k^{''})^{2}}{m} + i0} T^{MB}(k^{''},k,K;E)~.
\end{eqnarray}
Here $k$ and $k^{'}$ respectively correspond to ingoing and outcoming particle
momenta, $K$ to the centre-of-mass momentum and $E$ represents the kinetic
energy for the two atoms in the centre-of-mass frame. Furthermore, N represents
the average occupation number of a single particle state (of specified
momentum).

\subsection{Ladder Approximation to the Condensate-Excited State Interactions}

So far, we have only accounted for the occupation of states through which a
condensate-condensate collision proceeds, in terms of the many-body T-matrix.
The other important effect arising at finite temperatures, is the interaction
between condensed and excited atoms. Obviously, such an effect must also be
included in a NLSE describing the evolution of the condensate. To deal with
this effect, we must now also consider the rest of Eq. (8) that has been
ignored up to this point. To be more precise, we need to consider the effect of
the additional contribution
\begin{equation}
i \hbar \frac{dz_{n}}{dt} = ... + \sum_{ijk} \langle ni| \hat{V} | jk \rangle
\left[  2 \rho_{ji}z_{k} + \lambda_{ijk} \right]~.
\end{equation}
Careful observation of the first few terms in the equation of motion for
$\lambda_{ijk}$, Eq. (A7) suggests, by analogy to the arguments of Sec. II, that
it is indeed the suitable quantity for upgrading the actual (single-vertex)
interatomic potential for $2 \rho_{ji} z_{k}$ in Eq. (29) to the T-matrix level.
However, we note that the triplet $\lambda$ is not included (i.e.\ $\lambda=0$)
in the traditional mean field (i.e.\ HFB) theory. This shows that we need to go
beyond HFB in order to rigorously obtain a NLSE for the evolution of the
condensate mean field at non-zero temperatures, and we shall
discuss this below. An extension beyond HFB is,
strictly speaking, also essential for a consistent $T=0$ theory.
 
We shall now also adiabatically eliminate \cite{Foot4} the quantity
$\lambda_{ijk}$ appearing in Eq. (29). Consider initially the contribution
\begin{equation}
i \hbar \frac{d}{dt}  (\lambda_{ijk})  =  \hbar (\omega_{k} + \omega_{j} -
\omega_{i}) \lambda_{ijk} + \sum_{ms} \langle jk | \hat{V} | ms \rangle \left(
2 \rho_{mi} z_{s} + \lambda_{ims} \right)~.
\end{equation}
Thus, Eq. (29) becomes, to second order in the interaction potential
\begin{eqnarray}
i \hbar \frac{dz_{n}}{dt} & = & \cdots + \sum_{ims} \langle ni| \hat{V} | ms
\rangle 2 \rho_{mi} z_{s} + \sum_{ijk} \langle ni| \hat{V} | jk \rangle  \int \frac{dt^{'}}{i} e^{-i(\omega_{k} + \omega_{j} - \omega_{i})(t - t^{'})}  \nonumber
\\
&  & \times   \sum_{ms} \langle jk | \hat{V} | ms \rangle  2
\rho_{mi}(t^{'})z_{s}(t^{'})~.
\end{eqnarray}
As before, we assume that all mean value amplitudes $z_{i}(t)$ vary on a much
slower time-scale than the anomalous correlations. Furthermore, we limit our
discussion here to the slowly-evolving diagonal elements $\rho_{ii}^{0}
\delta_{mi}$ in a bare particle basis, for which $\rho_{ii}^{0}(t^{'}) =
\rho_{ii}^{0}(t)$. We thus obtain the desired upgrading of condensate-excited
state interactions to the two-body T-matrix, valid to second order in $V$, just
like in Eq. (17). Once again, we see that this expression can be
generalized to all orders in the interaction potential, by taking account of
the last term proportional to $\lambda_{ims}$ in the right-hand side of
Eq. (30).

Thus far, we have shown how to treat the condensate-condensate interactions to
the  bare-particle $\hat{T}^{MB}$ level and the condensate-excited state
contributions to $\hat{T}^{2B}$, so that the condensate mean field evolves
according to the equation
\begin{equation}
i \hbar \frac{dz_{n}}{dt} = \hbar \omega_{n} z_{n} + \sum_{ims} \langle ni|
\hat{T}^{MB} | ms \rangle  z_{i}^{*} z_{m} z_{s} + 2 \sum_{is} \langle ni|
\hat{T}^{2B} | is \rangle \rho_{ii}^{0}   z_{s}~.
\end{equation}
It is clear that a consistent finite temperature theory would require the
latter contribution to be also expressed in terms of $\hat{T}^{MB}$. The
subtle point that needs to be addressed here, is that such an expression will
contain terms of order $(\rho^{0})^{2}$; hence, we expect (and we shall indeed
confirm in our detailed treatment below) that in this situation, we should also
take account of the scattering of excited states into the condensate, which
modifies the evolution of the condensate mean value.

\subsection{Interactions Between Excited States}

In this section we shall discuss how to upgrade expression (32) to the
many-body T-matrix for excited states. On first inspection, this appears to be
quite straightforward, upon considering the first three terms of Eq. (A7), namely
the contribution of
\begin{eqnarray}
i \hbar \frac{d}{dt} (\lambda_{ijk}) & = &  \hbar (\omega_{j} + \omega_{k} -
\omega_{i}) \lambda_{ijk} + \sum_{rms} \langle jk | \hat{V} | ms \rangle \left(
2 \rho_{mi} z_{s} + \lambda_{ims} \right) \nonumber \\
& & + \left\{ \sum_{rms} \langle rk | \hat{V} | ms \rangle \rho_{jr} \left[ 2
\rho_{mi} z_{s} + \lambda_{ims} \right] + (k \leftrightarrow j) \right\}
\end{eqnarray}
to the equation for condensate evolution (29).

However, we wish to point out that maintaining merely this contribution for
$\lambda_{ijk}$ cannot be consistent. The reason for this is that the next term
in Eq. (A7), namely
\begin{equation}
i \hbar \frac{d}{dt}  (\lambda_{ijk})  =  \cdots  - \left\{ \sum_{rms} \langle
ms | \hat{V} | ir \rangle  \rho_{jm} \rho_{ks} z_{r} + (k \leftrightarrow j)
\right\}  \end{equation}
is an equally valid contribution to order $V^{2}$ in the expression for $d
z_{n}/dt$. Thus, a consistent approach must simultaneously take into account
both contributions (33) and (34) in adiabatically eliminating $\lambda$. The
interpretation of the contribution (33) is straightforward: it leads to the
renormalization of the condensate-excited state interaction to $\hat{T}^{MB}$.
At first sight, it appears hard to interpret the physical significance of (34).
To achieve this, we shall once again limit ourselves to the slowly-evolving
excited state contributions $\rho^{0}$. In this limit, maintaining both terms
(33) and (34), we obtain for the contribution of excited states to the
evolution of the mean value amplitude $z_{n}$ to second order in $V$,
\begin{eqnarray}
i \hbar \frac{dz_{n}}{dt} & = & \cdots + 2 \sum_{is} \langle ni | \hat{V} | is
\rangle \rho_{ii}^{0} z_{s} + 2 \sum_{ijks} \langle ni | \hat{V} | jk \rangle \int \frac{dt^{'}}{i} e^{- i (\omega_{k} + \omega_{j} - \omega_{i} - \omega_{s})(t -
t^{'})} \nonumber \\
& & \times \left[ \rho_{ii}^{0} + \rho_{ii}^{0} \rho_{jj}^{0} + \rho_{ii}^{0}
\rho_{kk}^{0} - \rho_{jj}^{0} \rho_{kk}^{0} \right](t^{'}) \langle jk | \hat{V}
| is \rangle z_{s}(t)~.
\end{eqnarray}

We can now easily identify the term in square brackets as
\begin{equation}
(\rho_{jj}^{0} + 1)(\rho_{kk}^{0} + 1) \rho_{ii}^{0} - \rho_{jj}^{0}
\rho_{kk}^{0} (\rho_{ii}^{0} + 1)~.
\end{equation}
This contribution is known to give the correct amplitude for the scattering of
quasiparticles, and can be used to predict condensate lifetimes. This
expression is also in agreement with the results of the functional integral
approach developed independently by one of us \cite{Nucl,Stoof_BECKin}.

Hence, we can now identify the negative contribution in (34) as the term which
ensures the correct factors for scattering into the condensate, due to the interaction of an excited state either with a condensed, or with another excited atom. Correct treatment of the factors of (36) to all
orders in the interaction potential should lead to an additional term in the
finite temperature NLSE, which should thus be expressable as
\begin{eqnarray}
i \hbar \frac{dz_{n}}{dt} = & & \omega_{n}z_{n} + \sum_{ims} \langle ni|
\hat{T}^{MB} | ms \rangle z_{i}^{*} z_{m} z_{s}
+ 2 \sum_{is} \langle ni| \hat{T}^{MB} | is \rangle \rho_{ii}^{0}
                                      z_{s} \nonumber \\
& &- 2 \sum_{ijks} \langle ni| \hat{T}^{MB} | jk \rangle \int \frac{dt^{'}}{i}
      e^{- i (\omega_{j} + \omega_{k} - \omega_{i} - \omega_{s})(t - t^{'})}
\left[ \rho_{jj}^{0} \rho_{kk}^{0}
                                                  \right](t^{'}) \nonumber \\
& & \times \langle jk | \hat{T}^{MB} | is \rangle z_{s}~.
\end{eqnarray}

Eq. (37) is depicted diagrammatically in Fig. 3. This equation is very important, as it includes the effect of kinetics due to
interactions of excited atoms. At the moment it appears that our formalism does
not generate the last many-body T-matrix, so that we merely obtain $\langle ms
| \hat{V} | ik \rangle$. However, consideration of the derivation
of the equations of appendix A  reminds us that correlations of products of
four fluctuation operators have only been treated by their mean field
contributions, i.e.\ in terms of products of averages of two operators (or
Wick's theorem, see Eq. (22) of reference \cite{Prou1}). This means that
our formalism does not yet fully include effects on the population of excited
states due to the interactions between two excited atoms. These can be taken
into account by explicitly deriving the equation of motion for $\langle
c^{\dag} c^{\dag} c c \rangle$ and carrying out the decomposition approximation
on higher order correlations. We have indeed shown \cite{Prou5} that, in the
limit of no condensation, such treatment gives rise to the well-known quantum
Boltzmann equation \cite{Cirac}. Thus, it appears reasonable that a full
treatment of these more complex correlations might solve the apparent
limitation of our formalism in terms of the derivation of Eq. (37). We shall return
to this issue in detail in a following paper \cite{Prou5}. However, since the last (kinetic) contribution arises due to the interaction of two excited atoms resulting in scattering into the condensate, this effect may actually be negligible in a sufficiently dilute and low temperature regime.

\section{Effect of Mean field on Collision Dynamics}

All discussion so far has been in terms of weakly-interacting atoms, for which
we can ignore the effect of the mean fields on the intermediate collisional
states. In particular, in adiabatically eliminating the anomalous averages
$\kappa_{jk}$ and $\lambda_{ijk}$, we have only considered certain of their
contributions appearing in the equations of motion of appendix A. These
contributions correspond to those giving rise to T-matrices for both
condensate-condensate and condensate-excited state interactions in the equation
governing the evolution of the condensate mean-field.

However, we know that the presence of mean fields (in terms of both condensate
as well as excited states) will modify or `dress' the intermediate collisional
states. In this section, we shall limit ourselves to the regime where all
effects of the triplets can be ignored ($\lambda=\gamma=0$). This will be shown
to correspond to the HFB description of the system in terms of actual
interatomic potentials, with the effect of mean fields during a collision
generating the quasiparticle (Bogoliubov) dressing. In Sec. V we will
discuss the validity of the simple GPE, as well as that of other conventional
dilute Bose gas theories and we shall argue that a consistent theory
necessitates an extension of the conventional mean field theory (i.e.\
consideration of the triplets in our present language).

\subsection{The limit where $\lambda=\gamma=0$}

Let us initially re-cast the equations of motion of appendix A in a simplified
manner in the limit $\lambda=\gamma=0$. In terms of the anomalous correlation
$\kappa_{jk}$, we note that Eq. (A6)  contains the additional (dressing)
contribution
\begin{equation}
i \hbar \frac{d \kappa_{kj}}{dt} = \cdots + \sum_{s} \left[ \eta_{ks}
\kappa_{sj}  + \kappa_{ks}  \eta_{sj}^{*}  \right]~,
\end{equation}
where $\eta$ is given by
\begin{equation}
\eta_{ks} = 2 \sum_{lt} \langle kl | \hat{V} | ts \rangle (z_{l}^{*} z_{t} +
\rho_{tl})~.
\end{equation}
From Eq. (A5), we also write down the equation of motion for $\rho_{ji}$.  In
Sec. III, we assumed $\kappa$ does not acquire a finite mean value
in-between collisions (i.e. it relaxes to zero) and also ignored the
off-diagonal $\rho$ elements. To deal with the conventional mean field
dressing, we shall now also allow for rapidly-varying off-diagonal
$\rho$-elements $\delta \rho_{ji}$, i.e we shall substitute for $\rho_{ji}$ the
expression \cite{Foot5} 
\begin{equation}
\rho_{ji} = \rho_{ii}^{0} \delta_{ji} + \delta \rho_{ji}~.
\end{equation}

We shall further assume that $d\rho_{ii}^{0}/dt \simeq 0$. In these limits, we
thus obtain the following set of equations for the rapid variations of $\delta
\rho$ and $\delta \kappa$ during a binary collisional process:

\begin{eqnarray}
i \hbar \frac{dz_{n}}{dt} & = & \hbar \omega_{n} z_{n} + \sum_{ijk} \langle ni|
\hat{V} | jk \rangle \left[ z_{i}^{*}z_{j}z_{k} +  2 \rho_{ii}^{0} \delta_{ij}
\right] \nonumber \\
& & + \sum_{ijk} \langle ni| \hat{V} | jk \rangle z_{i}^{*} \delta \kappa_{jk}
\nonumber \\
& & + 2  \sum_{ijk} \langle ni| \hat{V} | jk \rangle z_{k} \delta \rho_{ji}~,
\end{eqnarray}

\begin{eqnarray}
i \hbar \frac{d}{dt} (\delta \kappa_{kj}) & = & \hbar (\omega_{k} + \omega_{j})
\delta \kappa_{kj}  \nonumber \\
&  & + \sum_{s} \left[ \eta_{ks} \delta \kappa_{sj}  + \delta \kappa_{ks}
\eta_{sj}^{*}  \right] \\
& & + \left( 1 + \rho_{kk}^{0} + \rho_{jj}^{0} \right) \Delta_{kj} \\
& & + \sum_{s} \left[ \delta \rho_{ks} \Delta_{sj}  + \Delta_{ks}  \delta
\rho_{sj}^{*}  \right] ~,
\end{eqnarray}

\begin{eqnarray}
i \hbar \frac{d}{dt} (\delta \rho_{ji})  & = &  \hbar (\omega_{j} - \omega_{i})
\delta \rho_{ji} \nonumber \\
& & + \sum_{r} \left[ \eta_{jr} \delta \rho_{ri} - \delta \rho_{jr} \eta_{ri}
\right] \\
& & + \left( \rho_{ii}^{0} - \rho_{jj}^{0} \right) \eta_{ji} \\
&  & - \sum_{r} \left[\delta \kappa_{jr}  \Delta_{ri}^{*}  - \Delta_{jr}
\delta \kappa_{ri}^{*} \right]~.
\end{eqnarray}

The physical significance of the various contributions appearing in these equations will be analyzed in detail below. First, however, we would like to point out that the limit $\lambda=\gamma=0$ of the equations of
motion of appendix A corresponds to the time-dependent Hartree-Fock-Bogoliubov
(HFB) equations, written in terms of actual interatomic potentials. These HFB
equations can be alternatively generated by the following hamiltonian
\cite{Blaizot}
\begin{equation}
H_{Q} = \frac{1}{2}  \sum_{pq} \left\{ h_{pq} \left( \hat{c}_{p}^{\dag}
\hat{c}_{q} + \hat{c}_{q} \hat{c}_{p}^{\dag} \right) + \left( \Delta_{pq}
\hat{c}_{p}^{\dag} \hat{c}_{q}^{\dag} + \Delta_{pq}^{*} \hat{c}_{q} \hat{c}_{p}
\right) \right\}~,
\end{equation}
where
\begin{equation}
h_{pq} = \langle p | \hat{h} | q \rangle = \langle p | \hat{\Xi} | q \rangle +
\eta_{pq} = \langle p | \hat{\Xi} | q \rangle + 2 \sum_{lt} \langle pl |
\hat{V} | tq \rangle (z_{l}^{*} z_{t} + \rho_{tl})~.
\end{equation}
This operator takes account of all possible quadratic terms in bare particle
fluctuation operators $\hat{c}_{i}$. In the literature, $H_{Q}$ is termed the
quasiparticle hamiltonian \cite{Blaizot}, as it describes fully an assembly of
non-interacting quasiparticles. The dressing effect of the mean fields on the
bare-particle states resulting from this hamiltonian is conventionally referred
to as quasiparticle dressing, and we shall now discuss its consequences.

\subsubsection{Quasiparticle Dressing at Zero Temperature}

At $T \simeq 0$ there will be very few excited atoms in the assembly, so that
we can approximate $ \rho_{ii}^{0} \simeq 0$. We have already shown that the
term $i \hbar (d \kappa_{kj}/dt) \propto \Delta_{kj}$ of Eq. (43) corresponds to the
two-body T-matrix being generated in the interaction between condensed atoms.
The terms containing $\eta$ in both $\delta \rho$ and $\delta \kappa$ (i.e.
contributions (42) and (45)) give rise to the dressing of Fig. 4(a). This
represents the simplest dressing term due to the mean field of the condensate
in the intermediate collisional steps, and corresponds algebraically to
shifting and mixing of the intermediate states and frequencies.

From Eqs. (44) and (47), we see that there further exist more complex
contributions that depend on the coupling between $\delta \rho$ and $\delta
\kappa$. We expect these to correspond to the anomalous quasiparticle
(Bogoliubov) dressing, in much the same way creation and annihilation operators
 get coupled via the quasiparticle transformation
\begin{equation}
\hat{c}_{i} = \sum_{l} \left[ u_{il} \hat{b}_{l} - v_{il}^{*}
\hat{b}_{l}^{\dag} \right]
\end{equation}
where the $\hat{b}_{l}^{(\dag)}$ correspond to quasiparticle annihilation
(creation) operators.

Indeed, the combination of contributions (44) and (47) gives rise to
intermediate dressing of the form of Fig. 4(c). We stress that the coupling
between them implies that both diagrams in 4(c) are combined at any single
intermediate step. To understand the form of the coupling, we have illustrated
in Fig. 5 two typical diagrams in the evolution of the condensate mean field,
along with their dressed equivalents due to the $\delta \rho$-$\delta \kappa$
coupling.

\subsubsection{Quasiparticle Dressing at Nonzero Temperatures}

Observation of Eqs. (42)-(43) and (45)-(46) shows the following new
features arising at $T>0$, where $\rho_{ii}^{0}$ becoms non-negligible due to
thermally excited atoms. The first one, which we have already discussed quite
extensively, is the replacement of the two-body T-matrix by the many-body one,
due to enhanced condensate-condensate scattering  via occupied excited states.
Furthermore, the $\eta$ terms in (42) and (45) also allow modifications in the
intermediate collisional states due to the mean field of excited states (in
analogy to the condensate mean field of Fig. 4(a)), as shown in Fig. 4(b).

The final modification we need to consider at finite temperatures is that due
to the effect of the contribution (46). Adiabatic elimination of this contribution results in
the generation of the so-called bubble diagrams shown in Fig. 4(d). These
bubble diagrams can usually
be neglected in dilute systems, but have recently been shown to be rather
important near the critical temperature of the gas \cite{Stoof_Ren}.

The interplay between all dressings of Figs. 4 (a)-(d) in the many-body
T-matrix corresponds to what is conventionally refered to as the quasiparticle
dressing for the interactions between condensed atoms. Indeed, the consistent
combination of diagrams 4(a)-(c) with the T-matrix elements will generate the
well-known normal and anomalous self-energies $\hbar\Sigma_{11}$ and
$\hbar\Sigma_{12}$. However, we believe that the conventionally used finite
temperature mean field theory does not actually include all these effects in a
consistent fashion, as we shall argue below (Sec. V).

Following the above discussion on the dressing of interactions due to the
presence of mean fields, we shall now turn our attention to the analysis of
conventional Bose gas theories appearing in the literature.

\section{Conventional Theories and their Validity}

\subsection{The Gross-Pitaevskii Equation}

In section IV, we discussed the effect of mean fields on the intermediate
collisional states in the limit $\lambda=\gamma=0$. This effect means we should
use non-interacting quasiparticles, rather than bare particle states in our
theoretical treatment. This dressing is not taken into account in the GPE, as
was explicitly shown in its microscopic derivation in section II.  However,
this dressing can only be ignored in the limit of weak interactions, i.e. when
$n U_{0} \ll \hbar \omega$. In order to derive the GPE, we also have to ignore
the effects of the occupation of low-lying excitations during a collision
(many-body effects). These restrictions give us the sufficient condition for
the validity of the GPE, namely \begin{equation}
n U_{0} \, , \; k_{B}T \ll \hbar \omega~,
\end{equation}
where $\hbar \omega$ is a typical energy separation between trap levels.

Effectively, the GPE only treats condensate-condensate scattering in vacuum
(thus ignoring both quasiparticle dressing and many-body effects). In this
limit, the s-wave scattering length indeed contains all relevant collisional
information. It is conventionally argued that the corrections to the GPE due to
the presence of mean field are purely diluteness corrections. If the conditions of (51) are fulfilled the corrections are indeed small and this statement is close to the truth. However, even this statement is open to
critisism \cite{Prou5}. More generally, in the case of partial condensation ($T
\neq 0$), we maintain that the notion that diluteness is the issue is
misleading.

\subsection{Zero-Temperature Bogoliubov-de Gennes Equations}

As mentioned above, the GPE takes no account of the finite $T=0$ depletion of
the condensate, arising due to collisions between condensate atoms. This
depletion is indeed very small in the case of typical experimentally-studied
condensates, as has been shown explicitly, e.g. by Hutchinson et al.
\cite{Hutch}. One can, therefore, calculate the frequencies of the elementary
excitations, to good accuracy, by finding normal modes of the linearized GPE,
of the form
\begin{equation}
\Phi ({\bf r}, t) = e^{- i \mu t} \left[ \phi({\bf r}) + u({\bf r}) e^{- i
\omega t} +  v({\bf r}) e^{ i \omega t} \right]~.
\end{equation}
Here $\mu$ corresponds to the chemical potential of the undisturbed ground
state and the condensate wavefunction has been represented by the condensate
orbital $\phi({\bf r})$. In addition, $\omega$ labels the frequency of the
elementary excitations, whereas $u({\bf r})$ and $v({\bf r})$ correspond to
the spatially dependent coefficients of the condensate's linear response to
some driving field. This linear response approach has been discussed in
\cite{Lin_Resp}, \cite{NIST_Exc_Rev} where it has been shown that the
substitution (52) is completely equivalent to carrying out the Bogoliubov
transformation \cite{Bog} on the fluctuating part of the Bose field operator
$\hat{\Psi} - \langle \hat{\Psi} \rangle$, i.e. to the diagonalization of the
binary-interaction hamiltonian for the assembly.

Thus, one obtains a set of three static coupled equations for $\phi({\bf r})$,
$u({\bf r})$ and $v({\bf r})$, known as the Bogoliubov-de Gennes (BdG)
equations. These equations have been used to predict condensate shapes,
densities and the energies of elementary excitations at near zero temperatures.
The results of the BdG equations appear to be in excellent agreement
\cite{NIST_Exc_Comp} with experiments as discussed in the general
zero-temperature mean field theory review paper by Edwards et al.
\cite{NIST_Exc_Rev}.

\subsection{The `Essence' of Hartree-Fock-Bogoliubov}

We would like to point out that the equivalence between use of (52)
and the diagonalization discussed in \cite{Lin_Resp}, \cite{NIST_Exc_Rev} is
only true for a binary-interaction hamiltonian under the assumption $V({\bf r}
- {\bf r^{'}}) = U_{0} \delta ({\bf r} - {\bf r^{'}})$. In this paper, we have
argued that a $\delta$-function pseudopotential approximation can only be
imposed on an `upgraded' effective interaction potential expressable in terms
of the two-body T-matrix. This enormously complicates the discussion of the
validity of certain approximate mean field theories, and generates confusion as
to precisely what set of equations is implied by the term
`Hartree-Fock-Bogoliubov'.

We would argue, that the HFB equations are, in fact, defined in terms of actual
interatomic potentials. Thus, we define HFB as the set of coupled
$z$-$\rho$-$\kappa$ equations of appendix A (in which all
interaction terms are left in terms of symmetrized matrix elements), in the limit $\lambda=\gamma=0$. The time-dependent
HFB equations defined in this way, can thus be derived from the quasiparticle
hamiltonian (48) describing an assembly of non-interacting quasiparticles. We
have already argued in Sec. III (A), however, that such a theory cannot
consistently describe the interaction between an atom in the condensate with an
excited one. The rigorous way of doing this, is by consideration of the
$\lambda$ correlations. Bringing in the $\lambda$ correlations is equivalent to
allowing the quasiparticles to interact with each other.

We shall now discuss what people conventionally refer to as the finite temperature mean field, or HFB approximation.

\subsection{Finite Temperature Mean Field Theory and the Popov
Approximation}

The mean field treatment extensively discussed in the literature avoids dealing
with actual interatomic potentials. Instead, one conventionally expresses the
above finite temperature HFB equations in terms of a $\delta$-function
approximation. In the static case, these equations then take the form given
below. We would like to point out that these equations can be derived as a
special case of our microscopic formalism if one includes triplets. The
equations are

\begin{equation}
\left\{ -\frac{\hbar^2}{2M} \nabla^2 + V_{\rm trap}({\bf r}) - \mu +
U_0 \left[ N_0 \left| \psi({\bf r}) \right|^2 + 2 \tilde{n}({\bf r})
\right] \right\} \psi({\bf r}) + U_0 \tilde{m}({\bf r}) \psi^\ast({\bf
r}) = 0,
\end{equation}
\begin{equation}
\hat{\cal L} u_j({\bf r}) + U_0 \left[ N_0 \left\{ \psi({\bf r})
\right\}^2 + \tilde{m}({\bf r}) \right] v_j({\bf r}) = E_j u_j({\bf
r}),
\end{equation}
\begin{equation}
\hat{\cal L} v_j({\bf r}) + U_0 \left[ N_0 \left\{ \psi^\ast({\bf r})
\right\}^2 + \tilde{m}^\ast({\bf r}) \right] u_j({\bf r}) = -E_j
v_j({\bf r}),
\end{equation}
where we have defined the following quantities
\begin{equation}
\tilde{n}({\bf r}) = \sum_j \left\{ \left[ \left| u_j({\bf r})
\right|^2 + \left| v_j({\bf r}) \right|^2 \right] N_0 (E_j) + \left|
v_j({\bf r}) \right|^2 \right\}~,
\end{equation}
\begin{equation}
\tilde{m}({\bf r}) = \sum_j u_j({\bf r}) v_j^\ast({\bf r}) \left[ 2
N_0 (E_j) + 1 \right]~,
\end{equation}
and the operator
\begin{equation}
\hat{\cal L} = -\frac{\hbar^2}{2M} \nabla^2 + V_{\rm trap}({\bf r}) - \mu +
2 U_0 \left[ N_0 \left| \psi({\bf r}) \right|^2 + \tilde{n}({\bf r}) \right].
\end{equation}
Here, $N_0(E_j) = 1 / (e^{\beta E_j} - 1)$, and the above
quantities are normalized in the usual
manner \cite{Lin_Resp}. These equations are being used at the moment to study the finite temperature excitations of Bose-condensed gases \cite{Hutch,Dodd_Fin_Temp_Exc}. Since these equations do not use actual interatomic
potentials, we would rather not use the term HFB. In this paper, we shall call them the finite temperature Bogoliubov-de Gennes (BdG) equations, for want of a better name. The reason for making this distinction, arises from our view that the approximations made in obtaining Eqs. (53)-(55) actually contain more physics than the conventional mean field approximation. This can be seen from our microscopic approach: In Sec. III, we have explicitly shown that the condensate-excited state interactions can be
considered in terms of an effective T-matrix interaction only once the triplet
$\lambda$ is taken into account. We stress that this T-matrix
effective interaction is implicit in the above form of the finite temperature
BdG equations (53)-(55), since they have been written in terms of zero-range
potentials. It is clearly important to bear in mind that, when one includes the
$\lambda$ terms, one also brings in other effects that have been ignored in the
coupled BdG equations, such as further dressing of intermediate collisional
states. The finite temperature BdG equations, therefore, have some
inconsistencies built into them. How important these inconsistencies are in
practice remains to be seen.

Some discussion has been recently focused around the so-called `Popov'
approximation of these equations, which corresponds to setting  $\tilde{m}({\bf
r})=0$ in Eqs. (53)-(55) \cite{Griffin_HFB}. This
approximation has been used to compute the finite temperature excitations of a trapped Bose
gas \cite{Hutch,Dodd_Fin_Temp_Exc}. In these papers, the equations being solved are referred to as the coupled HFB-Popov
equations. We shall use the labelling Popov-BdG, since even the Popov approximation of these equations goes beyond the conventional mean field theory (due to the implicit inclusion of the $\lambda$ correlations). In our treatment, we have obtained these equations by
adiabatically eliminating both anomalous averages $\kappa$ (or equivalently
$\tilde{m}({\bf r})$) and $\lambda$ (not explicitly present in Eqs. (53)-(55)). Thus, the Popov limit appears to be one  way of obtaining a
consistent theory. Indeed, Griffin \cite{Griffin_HFB} has shown that such a
theory is gapless. However, using Eqs. (53)-(55) as they stand, with a nonzero value of
$\tilde{m}({\bf r})$ is clearly inconsistent, because we have shown that it is precisely the adiabatic elimination of $\tilde{m}({\bf r})$ which leads to the replacement of the interatomic potential by an effective interaction $U_0$.

Let us now discuss the validity of the Popov approximation. The BdG equations
have been written in terms of the quasiparticle coherence factors $u_j({\bf
r})$ and $v_j({\bf r})$. This shows explicitly that they include the effect of
the condensate mean field on the initial and final collisional states, which modifies the bare particle states into
non-interacting Bogoliubov quasiparticles. However, the analysis of Sec. IV
shows that in this limit, we must also consider other effects, like the
dressing of intermediate collisional states and the occupation of these
excitations. These many-body effects should be fully included in a consistent
theory of condensate-condensate and condensate-excited state collisions. The
Popov-BdG theory does not take account of these effects, and we would thus
expect it to deviate from the actual description of the system. Indeed, a
basis-set simulation of excitation frequencies using the Popov-BdG equations
¦\cite{Dodd_Fin_Temp_Exc}, has revealed large (and qualitative) differences from experimental
data for temperatures above $0.6 T_{c}$. We are currently working on
determining a more appropriate set of equations --- similar to what we have termed finite temperature BdG equations --- that can be straightforwardly computed.

Our treatment shows that many-body effects are not the only effects that are
not included in the Popov-BdG description, as discussed below.

\section{Effects of Mean Fields Beyond Quasiparticle Dressing}

The equations of appendix A, indicate that the dressing discussed in Sec. IV
is not the only effect the mean fields have during a collision. The dressing of
Sec. IV arises from the quasiparticle hamiltonian (48), which replaces the
interacting (bare) atoms by non-interacting quasiparticles. However, we believe
the theory should allow the quasiparticles to interact weakly with each other.
These interactions will lead to more complex dressing of the intermediate
collisional states. However, this is not the only reason for going beyond a
quasiparticle description. We have extensively shown in this paper that, to
obtain the evolution of the condensate mean field at finite temperatures, we
need to consider the triplet $\lambda_{ijk}$ in Eq. (8). However, once including
this, there appears to be no valid argument for neglecting the effects of the
triplets in the remaining equations of appendix A. For example, Eq. (A6) shows that
when adiabatically eliminating $\kappa_{jk}$, we will also have to worry about
triplet effects.

We thus believe the triplets have two effects. The reader is by now familiar
with the first one, which is the necessity of triplets for deriving finite
temperature equations for condensate evolution. Secondly, to extend this
argument even further, we also think that some of the triplets appearing in
Eq. (A6) are actually needed in order to consistently combine the dressing diagrams
of Fig. 4. We believe that a detailed study of the effects of the triplets
may shed some light into why the dressed many-body T-matrix appears to go to
zero as $T \rightarrow 0$ \cite{Stoof_Var,Stoof_Ren}. This behavior implies a
vanishing interaction (in the nonlinear Schr\"{o}dinger equation) for
collisions between two condensate atoms, which cannot be correct by itself, since we know
that it is precisely these collisions which lead to the finite zero-temperature
depletion of the condensate. We hope to discuss this, and related issues, in
the future.

\section{Conclusions}

In this paper we have carried out an in-depth analysis of our microscopic
description of the behavior of Bose-condensed systems at finite
temperatures. The  equations of \cite{Prou1} have been re-cast in Appendix A,
for the convenience of the reader.

After a general introduction into our microscopic approach, we used the set of
time-dependent Hartree-Fock-Bogoliubov (HFB) equations based on actual
interatomic potentials (i.e.\ without making any assumptions about them), to
derive a nonlinear Schr\"{o}dinger equation (16) for the description of the
condensate mean field evolution at temperatures close to zero; in this limit,
we can, to a good approximation, neglect the presence of excitations. We showed
explicitly how to `upgrade' the interatomic interaction potential (5) to the
bare-particle two-body T-matrix (18) which includes all repeated binary
collisional processes in vacuum. This is made possible by adiabatic elimination
of the anomalous correlation $\langle c c \rangle$  which is not retained in
the conventional derivation of the GPE.

Furthermore, we explicitly showed in what limits the bare-particle NLSE (16)
reduces to the GPE (7), thus providing a clear microscopic derivation. We then
generalized this NLSE to finite temperatures by including the effects of
excited states. This results in a variety of new features: Firstly, the
occupation of excited states during a collision leads to the replacement of the
two-body  T-matrix for bare particles by the many-body one. Secondly, we must
now also consider the effect of condensate-excited state interactions in the
evolution of the condensate mean field. In order to do this consistently, we
have shown the necessity to extend the conventional mean-field theory by
explicitly including the triplet $\langle c^{\dag} c c \rangle$, which upgrades
this interaction to the T-matrix. We have thus argued how to obtain a finite
temperature equation for the evolution of the condensate mean-field, which
includes kinetic contributions due to collisions between excited atoms. All
such treatment is true in the limit of weak interactions, i.e. when $n U_{0}
\ll \hbar \omega$.

When this condition does not hold, we must take account of the effect the mean
fields have on the intermediate collisional states. This leads to dressing of the
states accessed during a binary collision. The presence of such dressing
contributions   makes it extremely difficult to adiabatically eliminate the
anomalous correlations $\langle c c \rangle$ and $ \langle c^{\dag} c c
\rangle$ in favour of a consistent equation valid in all limits. In fact, we
point out that the triplet $\langle c^{\dag} c c \rangle$ must be rigorously
dealt with in a consistent $T=0$ theory, due to the finite (albeit negligible)
condensate depletion. Our microscopic approach enables us to discuss the
conventional mean-field theories currently used for describing the evolution of
the condensate. In first instance, we have argued that the Gross-Pitaevskii
equation is strictly only valid in the regime $n U_{0}, \; k_{B} T \ll \hbar
\omega$. Furthermore, we have argued that the conventionally used finite
temperature Bogoliubov-de Gennes equations contain some inconsistencies if they
are used beyond the Popov approximation. We have also argued that even the
Popov limit of these equations (which is generally believed to form a consistent theory \cite{Griffin_HFB}) may fail due to the neglect of many-body effects
on the collisions in the gas. Indeed, recent simulations
\cite{Dodd_Fin_Temp_Exc} have shown huge discrepancies with the experiments at
temperatures beyond $0.6 T_{c}$.

The analysis carried out in this paper shows precisely how hard it is to obtain
a consistent mean field theory for the description of partially Bose-condensed
systems. An alternative description that may not face the same difficulties is
based on Popov's approach of describing the homogeneous Bose gas in terms of an
effective condensate density and phase. Such an approach has already been
discussed by Ilinski and Stepanenko \cite{Ilinski}, and we hope more
discussions in this area will appear in the future.

\acknowledgements
The authors are grateful for helpful discussions with I. Cirac, M. Edwards, M.
Bijlsma, S. Morgan, C. W. Clark, R. J. Dodd, and P. Julienne. N.
P. Proukakis would like to thank the Alexander S. Onassis Public Benefit
Foundation (Greece) for financial support, whereas K. Burnett would like to
acknowledge funding from the European Community, under the TMR network
programme.

\appendix \section{Generalized Mean Field Equations}

In this appendix we give the full set of self-consistent mean field equations
derived in \cite{Prou1}, subject to a decoupling approximation of correlations
of four and five fluctuation operators. We have re-expressed these equations in
a form that allows us to identify the physical importance of all the individual
contributions. We remind the reader of the following definitions
\begin{equation}
\rho_{ji} = \langle \hat{c}_{i}^{\dag} \hat{c}_{j} \rangle~, \hspace{0.8cm}
\kappa_{jk} = \langle \hat{c}_{k} \hat{c}_{j} \rangle~, \hspace{0.8cm}
\lambda_{ijk} = \langle \hat{c}_{i}^{\dag} \hat{c}_{j} \hat{c}_{k} \rangle~,
\hspace{0.8cm} \gamma_{ijk} = \langle \hat{c}_{i} \hat{c}_{j} \hat{c}_{k}
\rangle~, \nonumber
\end{equation}
\begin{equation}
\eta_{jr} =  2 \sum_{lt} \langle jl | \hat{V} | tr \rangle (z_{l}^{*} z_{t} +
\rho_{tl})~, \nonumber
\end{equation}
\begin{equation}
\Delta_{kj} = \sum_{ms} \langle kj| \hat{V} | ms \rangle \left[ z_{m} z_{s} +
\kappa_{ms}  \right]~. \nonumber
\end{equation}
In writing down the equations of motion, we have assumed we are working in a
bare particle basis, where the operator $\hat{\Xi} = - (\hbar^{2} \nabla^{2})/(2m) + V_{trap}$ is diagonal, i.e $\langle k | \hat{\Xi} | n \rangle = \hbar
\omega_{n} \delta_{nk}$. Furthermore, the equations given below are expressed
in terms of the actual interatomic potential experienced between two atoms at
each collisional vertex. We thus obtain the following set of equations

\begin{equation}
i \hbar \frac{dz_{n}}{dt}  = \hbar \omega_{n} z_{n} + \sum_{ijk} \langle ni|
\hat{V} | jk \rangle \left[ z_{i}^{*}z_{j}z_{k} + \kappa_{jk}z_{i}^{*} + 2
\rho_{ji}z_{k} + \lambda_{ijk} \right]~,
\end{equation}

\begin{eqnarray}
i \hbar \frac{d \rho_{ji} }{dt} & = &  \hbar (\omega_{j} - \omega_{i})
\rho_{ji} \nonumber \\
& & + \sum_{r} \left[ \eta_{jr} \rho_{ri} - \rho_{jr} \eta_{ri} \right]
\nonumber \\
&  & - \sum_{r} \left[\kappa_{jr}  \Delta_{ri}^{*}  - \Delta_{jr}
\kappa_{ri}^{*} \right] \nonumber \\
 & &  +\sum_{rms} \langle jr| \hat{V} | ms \rangle \left( 2 \lambda_{mri}^{*}
z_{s} + \lambda_{ims} z_{r}^{*} \right) \nonumber \\
&  & - \sum_{rms} \langle ms| \hat{V} | ri \rangle
\left( 2 \lambda_{mrj} z_{s}^{*} + \lambda_{jms}^{*} z_{r} \right)~,
\end{eqnarray}

\begin{eqnarray}
i \hbar \frac{d \kappa_{kj}}{dt} & = & \hbar (\omega_{k} + \omega_{j})
\kappa_{kj}  \nonumber \\
& & + \Delta_{kj} + \sum_{s} \left[ \rho_{ks} \Delta_{sj}  + \Delta_{ks}
\rho_{sj}^{*}  \right] \nonumber \\
&  & + \sum_{s} \left[ \eta_{ks} \kappa_{sj}  + \kappa_{ks}  \eta_{sj}^{*}
\right] \nonumber \\
& &  + \sum_{rms} \left\{ \langle kr| \hat{V} | ms \rangle \left[ 2
\lambda_{rsj} z_{m} + \gamma_{msj} z_{r}^{*} \right] + (k \leftrightarrow j)
\right\} ~,
\end{eqnarray}

\begin{eqnarray}
i \hbar \frac{d}{dt}  (\lambda_{ijk})  & = & \hbar (\omega_{k} + \omega_{j} -
\omega_{i}) \lambda_{ijk} \nonumber \\
& & + \sum_{ms} \langle jk | \hat{V} | ms \rangle \left( 2 \rho_{mi} z_{s} +
\lambda_{ims} \right) \nonumber \\
& & +  \sum_{rms} \left\{ \langle kr | \hat{V} | ms \rangle \rho_{jr} \left[ 2
\rho_{mi} z_{s} + \lambda_{ims} \right] + (k \leftrightarrow j) \right\}
\nonumber \\
& & -  \sum_{rms} \left\{ \langle ms | \hat{V} | ir \rangle  \rho_{jm}
\rho_{ks} z_{r} + (k \leftrightarrow j) \right\} \nonumber \\
& & + \sum_{s} \left\{ \lambda_{ijs} \eta^{*}_{sk} + \lambda_{jis}^{*}
\Delta_{sk} + (k \leftrightarrow j) \right\}  \nonumber \\
& & - \sum_{s} \left[ \eta_{is}^{*} \lambda_{sjk} - \Delta_{js}^{*}
\gamma_{sik} \right] \nonumber \\
& & + \sum_{rms} \left\{ \langle kr | \hat{V} | ms \rangle 2 \rho_{mi}
(\kappa_{js} z_{r}^{*} + \lambda_{rjs}) +  (k \leftrightarrow j) \right\}
\nonumber \\
& & - \sum_{rms}  \langle ms | \hat{V} | ir \rangle \left\{ \kappa_{rk} (2
\rho_{jm} z_{s}^{*} + \lambda_{jms}^{*}) + (k \leftrightarrow j) \right\}
\nonumber \\
& & + \sum_{rms} \left\{ \langle kr | \hat{V} | ms \rangle \left[ 2 \kappa_{mj}
(\kappa_{ir}^{*} z_{s} + \lambda_{sir}^{*}) + \kappa_{ir}^{*} \gamma_{smj}
\right]   + (k \leftrightarrow j) \right\}\nonumber \\
& & - \sum_{rms}  \langle ms | \hat{V} | ir \rangle \left\{ 2 \rho_{jm}
\lambda_{skr} +  (k \leftrightarrow j) \right\}~,
\end{eqnarray}

\begin{eqnarray}
i \hbar \frac{d}{dt}  (\gamma_{ijk})  & = &  \hbar (\omega_{i} + \omega_{j} +
\omega_{k}) \gamma_{ijk} \nonumber \\
& & + \sum_{ms} \langle ij | \hat{V} | ms \rangle \left( 2 \kappa_{km} z_{s} +
\gamma_{kms} \right) \nonumber \\
& & +  \sum_{rms} \left\{ \langle ir | \hat{V} | ms \rangle \rho_{jr} \left[ 2
\kappa_{km} z_{s} + \gamma_{kms} \right] + (i \leftrightarrow j) \right\}
\nonumber \\
& & + \sum_{rms}  \langle ir | \hat{V} | ms \rangle \left\{ \kappa_{mk} \left[
\kappa_{js} z_{r}^{*} + 2 \lambda_{rjs} \right] + (k \leftrightarrow j)
\right\} \nonumber \\
& & + \sum_{s} \eta_{is} \gamma_{sjk} \nonumber \\
& & + (i \leftrightarrow j \leftrightarrow k) ~.
\end{eqnarray}
(The term $ + (i \leftrightarrow j \leftrightarrow k)$ in the last equation indicates summation of {\em all} terms appearing in (A8), under cyclic rotation of the indices $i$, $j$, and $k$ --- except, of course, the first `free evolution' contribution.)

\section{Condensate Evolution to Second Order in the Interaction Potential}

In this appendix we give a systematic categorization of all contributions to
the condensate mean field evolution, to second order in the interaction
potential $V$. We shall carry this out by making use of the decomposition (40)
of the single-particle correlation $\rho$ into a slowly-varying diagonal
($\rho^{0}$) and a rapidly-evolving off-diagonal element ($\delta \rho$). The
implicit assumption made here, is that the atoms are weakly-interacting, so
that the bare particle basis may be used, to a good approximation, for the
description of the system. Furthermore, we have already argued in Sec. III
(C) that the equations of the appendix do not  take account of the evolution of
excited states due to quantum Boltzmann-type effects, which can be included by
suitable treatment of the $\langle c^{\dag} c^{\dag} c c \rangle$ correlations.
We shall therefore assume that during a collision  $d \rho_{ii}^{0}/dt \simeq
0$, which gives for the condensate mean field evolution to $V^{2}$, the
expression
\begin{eqnarray}
i \hbar \frac{dz_{n}}{dt} & = & \hbar \omega_{n} z_{n} + \sum_{ims} \langle ni|
\hat{V} | ms \rangle  z_{i}^{*} z_{m} z_{s} \\
& & + 2 \sum_{is} \langle ni| \hat{V} |
is \rangle \rho_{ii}^{0}   z_{s} \\
& & + \sum_{ijk} \langle ni| \hat{V} | jk \rangle \sum_{ms} \int \frac{dt^{'}}{i}
e^{-i(\omega_{k} + \omega_{j} - \omega_{m} - \omega_{s})(t - t^{'})} \nonumber
\\
& & \times ( 1 + \rho_{jj}^{0}(t^{'}) + \rho_{kk}^{0}(t^{'}) ) \langle jk |
\hat{V} | ms \rangle z_{i}^{*}(t) z_{m}(t) z_{s}(t)  \\
& & + \sum_{ijk} \langle ni| \hat{V} | jk \rangle \sum_{s} \int \frac{dt^{'}}{i}
e^{-i(\omega_{k} + \omega_{j} - \omega_{i} - \omega_{s})(t - t^{'})} \nonumber
\\
& & \times ( 1 + \rho_{jj}^{0}(t^{'}) + \rho_{kk}^{0}(t^{'}) ) \langle jk |
\hat{V} | is \rangle \rho_{ii}^{0}(t) z_{s}(t)  \\
& & - \sum_{ijk} \langle ni| \hat{V} | jk \rangle \sum_{s} \int \frac{dt^{'}}{i}
e^{-i(\omega_{k} + \omega_{j} - \omega_{i} - \omega_{s})(t - t^{'})} \nonumber
\\
& & \times 2 \rho_{jj}^{0}(t^{'}) \rho_{kk}^{0}(t^{'})  \langle jk | \hat{V} |
is \rangle z_{s}(t)   \\
& & + 4 \sum_{ijk} \langle ni| \hat{V} | jk \rangle z_{k}(t)  \sum_{ms}
\int \frac{dt^{'}}{i} e^{-i(\omega_{i} + \omega_{m} - \omega_{j} - \omega_{s})(t -
t^{'})}  \nonumber \\
&  & \times  \left[ \rho_{ii}^{0}(t^{'}) - \rho_{jj}^{0}(t^{'}) \right]
\langle jm | \hat{V} | is \rangle  \left[ z_{m}^{*}(t) z_{s}(t) +
\rho_{ss}^{0}(t) \delta_{sm} \right]~,
\end{eqnarray}
where $t^{'}$ acquires values within the range $-\infty$ to $t$. This equation has been diagrammatically depicted in Fig. 6, where we also give the contribution factors of each diagram.

The terms in (B1)-(B2) represent the free evolution of the condensed particle, and
the interaction of one condensed atom, either with another condensed atom (Fig. (a)), or
with an excited one (Fig (b)). (B3) and (B4) give the lowest order corrections of the
actual interatomic potential, due to its replacement by the many-body T-matrix,
in both cases of condensate-condensate (Fig. 6(c)) and
condensate-excited state interactions (Fig. 6(d)). The contribution (B5) --- Fig. 6(e) --- arises due to the interactions between two excited atoms and ensures the
correct scattering amplitude factors for condensate evolution, as already explained in Sec. III (C).
Finally, (B6) corresponds to the bubble diagrams of the many-body formalism, which are shown in Fig. 6(f). 

At this point, we should really comment on the consistency of the above
equation for condensate evolution. In obtaining this equation, we have not
dealt with all quantities in the same manner. A fully consistent treatment
would be to assign a slowly-varying and a rapidly evolving part to each of the
quantities $\rho$, $z$ and $\kappa$ (which would also generate two parts for
both $h$ and $\Delta$), as well as the triplets $\lambda$ and $\gamma$ and then
deal with them as a closed system. We shall return to this issue in a
forthcoming publication \cite{Prou5}.

\clearpage

\section{Figure Captions}

Figure 1: Fig. (a) depicts the contribution to the evolution of the condensate mean field (level n) due to the instantaneous interaction of two condensate
atoms (labelled by $j$ and $k$). In this, and all subsequent figures, the curly line
represents the vertex between two interacting atoms located at ${\bf r}$ and
${\bf r^{'}}$ respectively, and time runs vertically upwards. We have explicitly drawn the two contributions to the evolution of the condensate mean
field due to the physical non-symmetrized matrix elements of Eq. (6)
separately. Each of these diagrams is associated with a factor of
$(\frac{1}{2})$ into our final equation for the evolution of $z_{n}$, as
defined by Eq. (5). In our notation, continuous lines with
one `free' end-point represent condensate particles, and have a factor $z$ associated with them (the dashed condensate line gives no such contribution, as it is the one whose evolution we are monitoring in this paper). Fig. 1(b) shows the cumulative effect of repeated binary interactions of this type, or ladder diagrams, in which an arbitrary
number $p$ of loops ($p \ge 0$) may be present. Here, the final scattering into
(condensate) states $n$ and $i$ has been mediated by other excited states
(given by the vertical lines with arrows).

Figure 2: Fig. (a) depicts diagrammatically the integral definition of the
bare-particle two-body T-matrix, as defined by Eq. (18). In Figs.
(b)-(c) we have illustrated the definition of the many-body T-matrix according
to (26). The new feature introduced in (b) is that one now takes account of the
occupation of the intermediate states through which the collision proceeds.
Thus, the thick arrows indicate that the collisions actually occur in an
atomic medium, as opposed to the vacuum. In particular, Fig. (c) indicates explicitly how this
excited state occupation is taken into account according to the many-body
factor $(1 + \rho_{jj}^{0} + \rho_{kk}^{0})$, with the $\rho^{0}$ term
indicated by the double arrow.

Figure 3: This figure represents diagrammatically the evolution of the
condensate mean value amplitude $z_{n}$ as given by Eq. (37). We have explicitly illustrated diagrams corresponding to direct and
exchange terms, as well as the associated contribution pre-factors. These
factors have been suppressed in our mathematical analysis, by means of the
definition (5) of a symmetrized matrix element. The factor $\rho^{0}$ indicates
the population of excited states.

Figure 4: This figure illustrates all different quasiparticle dressing effects
on the unoccupied intermediate propagators of the two-body T-matrix (Fig.
2(a)). Such dressing during an atomic collision arises due to the effect of (a)
the condensate mean field, or (b) the mean field of thermally excited states.
Additionally, the $\delta \rho$-$\delta \kappa$ coupling generates the
anomalous terms of Fig. (c), corresponding to the creation, or annihilation
of two condensed atoms, in favour of excited ones; these contributions do not
appear separately, but are complementary of each other. Fig. (d) indicates
the extra dressing due to the bubble diagrams (46) which are associated with a
factor ($\rho_{ii}^{0} -\rho_{jj}^{0}$). We note that all above dressing
effects could also be very straigthforwardly included in the many-body loops of
Fig. 2(b)-(c).

Figure 5: This figure illustrates the cooperative action of the $\delta
\rho$-$\delta \kappa$ coupling represented in Fig. 4(c), by means of two
typical terms in the evolution of the condensate mean field, and their
corresponding dressing due to this coupling: (a) shows the effect on the
scattering between a condensed and an excited atom, whereas (b) corresponds to
the dressing on a typical two-body ladder diagram.

Figure 6: This figure illustrates the evolution of the condensate mean field to second order in the interaction potential. Figs. (a)-(d) show the many-body T-matrix terms in the interactions (to order $V^{2}$) between two condensed atoms (Figs. (a),(c)), or one condensed atom with an excited one (Figs. (b),(d)). Fig. 6(e) shows the contribution to condensate evolution, due to the interaction of two excited atoms. Fig. 6(f) corresponds to the bubble diagrams, which have a factor $\left( \rho_{ii}^{0} - \rho_{jj}^{0} \right)$ associated with them (indicated by the hollow arrows). Continuous lines with one `free' end-point have a factor $z$ associated with them, and the thick arrows due to excited state occupation have been defined in Fig. 2(c).


\begin{references}

\bibitem{JILA} M.H. Anderson, J.R. Enscher, M.R. Matthews, C.E. Wieman and E.A.
Cornell, Science {\bf 269}, 198 (1995).
\bibitem{MIT} K.B. Davis, M.-O. Mewes, M.R. Andrews, N.J. van Druten, D.S.
Durfee, D.M. Kurn and W. Ketterle, Phys. Rev. Lett. {\bf 75}, 3969 (1995).
\bibitem{Rice} C.C. Bradley, C.A. Sackett, J.J. Tollett and R.G. Hulet, Phys. Rev. Lett. {\bf 75}, 1687 (1995); C. C. Bradley, C. A. Sackett and R. G. Hulet, Phys. Rev. Lett. {\bf 78}, 985 (1997).
\bibitem{NLSE} E. P. Gross, J. Math. Phys. {\bf 4} 195 (1963) and 
L. P. Pitaevskii, Sov. Phys. JETP {\bf 13}, 451 (1961).
\bibitem{keith} M. Edwards and K. Burnett, Phys. Rev. A {\bf 51}, 1382 (1995).
\bibitem{Pete} P.A. Ruprecht, M.J. Holland, K. Burnett and M. Edwards, Phys.
Rev. A {\bf 51}, 4704 (1995).
\bibitem{Ed} M. Edwards, R. J. Dodd, C. W. Clark, P.A. Ruprecht, and K.
Burnett, Phys. Rev. A {\bf 53}, 1950 (1996).
\bibitem{Holland} M. J. Holland and J. Cooper, Phys. Rev. A {\bf 53}, R1954 (1996).
\bibitem{Baym} G. Baym and C. Pethick, Phys. Rev. Lett. {\bf 76}, 6 (1996).
\bibitem{NIST_Exc_Comp} M. Edwards, P. A. Ruprecht, K. Burnett, R. J. Dodd and
C. W. Clark, Phys. Rev. Lett. {\bf 77}, 6 (1996).
\bibitem{Lin_Resp} P.A. Ruprecht, M. Edwards, K. Burnett and C.W. Clark, Phys.
Rev. A {\bf 54}, 4178 (1996).
\bibitem{Fett} A.L. Fetter, Phys. Rev A {\bf 53}, 4245 (1996).
\bibitem{Dalfovo} F. Dalfovo and S. Stringari,  Phys. Rev. A {\bf 53}, 2477
(1996).
\bibitem{NIST_Vort_1} R. J. Dodd, K. Burnett, M. Edwards and C. W. Clark,
preprint (1996).
\bibitem{NIST_Exc_Rev} M. Edwards, R. J. Dodd, C. W. Clark and K. Burnett, J.
Res. Nat. Inst. Stand. Tech. {\bf 101}, 553 (1996).
\bibitem{Dal_NIST} F. Dalfovo, L. Pitaevskii and S. Stringari, J.
Res. Nat. Inst. Stand. Tech. {\bf 101}, 537 (1996).
\bibitem{Exc} The first experiments on measuring the excitations of a
Bose-Einstein condensate have been reported in: D. S. Jin, J. R. Ensher, M. R.
Matthews, C. E. Wieman and E. Cornell, Phys. Rev. Lett. {\bf 77}, 420 (1996)
and M. -O. Mewes, M. R. Andrews, N. J. van Druten, D. M. Murn, D. S. Durfee, C.
G. Townsend and W. Ketterle, Phys. Rev. Lett. {\bf 77}, 988 (1996). Recent
temperature dependent damping measurements have also been reported in D. S. Jin, M. R. Matthews, J. R. Ensher, C. E. Wieman and E. A. Cornell, Phys. Rev. Lett. {\bf 78}, 764 (1997).
\bibitem{Stoof_Var} M. Bijlsma and H.T.C. Stoof, Phys. Rev. A {\bf 55}, 498
(1997).
\bibitem{Nucl} H.T.C. Stoof, Phys. Rev. A {\bf 45}, 8398 (1992).
\bibitem{Prou1} N.P. Proukakis and K. Burnett, J. Res. Natl. Inst. Stand.
Technol. {\bf 101}, 457 (1996).
\bibitem{Foot2} An alternative approach using Green's functions has been independently carried out by Griffin \cite{Griffin_HFB}. In Griffin's approach, however, the HFB
equations are considered in the limit of $\delta$-function pseudopotentials, as
will be discussed in more detail in section V.
\bibitem{Griffin_HFB} A. Griffin, Phys. Rev. B {\bf 53}, 9341 (1996).
\bibitem{HFB1} D.A. Huse and E.D. Siggia, J. of Low Temp. Phys. {\bf 46}, 137
(1982).
\bibitem{Dorre} P. D\"{o}rre, H. Haug and D.B. Tran Thoai, J. of Low Temp.
Phys. {\bf 35}, 465 (1979).
\bibitem{Kobe} D. H. Kobe, Annals of Phys. {\bf 47}, 15 (1968).
\bibitem{Blaizot} J. P. Blaizot and G. Ripka, Quantum Theory of Finite Systems,
MIT Press (1986).
\bibitem{Stoof_Attr} H. T. C. Stoof, Phys. Rev. A {\bf 49}, 3824 (1994).
\bibitem{Laloe} F. Lal\"{o}e in Bose-Einstein Condensation, edited by A.
Griffin, D.W. Snoke and S. Stringari, Cambridge University Press (1995).
\bibitem{Prou_RoyalSoc} N. P. Proukakis and K. Burnett, Phil. Trans. R. Soc.
Lond. A {\bf 355} (1997).
\bibitem{Foot3} We point out that this
distinction between the matrix elements $\langle ni| \hat{V} |  jk \rangle$ and $\left( ni| \hat{V} |  jk \right)$ was not explicitly analyzed in \cite{Prou1}, although all
equations of motion appearing there (i.e. all equations after (17) in
\cite{Prou1})¦made use of this symmetrized matrix element of the interaction
potential. We note that the use of the symmetrized matrix element greatly
simplifies all subsequent expressions, and has therefore also been  adopted in
this paper.
\bibitem{Fermi} E. Fermi, Riverca Sci. {\bf 7}, 13 (1936).
\bibitem{Huang_Yang} K. Huang and C. N. Yang, Phys Rev {\bf 105}, 767 (1957).
\bibitem{Nozieres} P. Nozi\`{e}res and D. Pines, {\em The Theory of Quantum
Liquids, Volume II}, Addison-Wesley (1990).
\bibitem{Huang_NIST} K. Huang and P. Tommasini,  J. Res. Nat. Inst. Stand.
Tech. {\bf 101}, 435 (1996).
\bibitem{Huang_Book} K. Huang, {\em Statistical Mechanics}, 2nd ed., John Wiley
and Sons (1987).
\bibitem{You} M. Lewenstein and L. You, Phys. Rev. A {\bf 53}, 909 (1996).
\bibitem{Prou5} N. P. Proukakis, M. Rusch and K. Burnett, in preparation
(1997).
\bibitem{Stoof_NIST} H.T.C. Stoof, M. Bijlsma and M. Houbiers, J. Res. Natl.
Inst. Stand. Technol. {\bf 101}, 443 (1996).
\bibitem{Foot4} Here we are assuming that
the anomalous correlation $\lambda_{ijk}$ acquires no equilibrium value in the
assembly, just as was done in section II for $\kappa_{jk}$.
\bibitem{Stoof_BECKin} H. T. C. Stoof, Phys. Rev. Lett. {\bf 78}, 768 (1997).
\bibitem{Cirac} This was first suggested to us in terms of our formalism by
Ignacio Cirac.
\bibitem{Foot5} At this point, we note that we are interested in the
evolution of the condensate mean field. If we were dealing with dressing on the
evolution of excited bare particle states, for example, we would also need to
consider a split $z \rightarrow z^{0} + \delta z$; a fully self-consistent
approach would further necessitate $\kappa \rightarrow \kappa^{0} + \delta
\kappa$ (implying that we no longer need to assume that $\kappa(t_{0})
\rightarrow 0$ as $t_{0} \rightarrow - \infty$), as well as $\eta \rightarrow
\eta^{0} + \delta \eta$ and $\Delta \rightarrow \Delta^{0} + \delta \Delta$. We
shall return to this exact linearisation procedure in a future publication
dealing with the damping of quasiparticles \cite{Prou5}.
\bibitem{Stoof_Ren} M. Bijlsma and H. T. C. Stoof, Phys. Rev. A {\bf 54}, 5085
(1996).

\bibitem{Bog} N. N. Bogoliubov, J. Phys. U.S.S.R. {\bf 11}, 23 (1947).
\bibitem{Hutch} D. A. Hutchinson, E. Zaremba and A. Griffin, Finite
temperature excitations of a trapped Bose gas, preprint (1996).
\bibitem{Dodd_Fin_Temp_Exc} R. J. Dodd, M. Edwards, C. W. Clark and K. Burnett,
subm. to Phys. Rev. Lett. (1997).
\bibitem{NIST_Attr} R. J. Dodd, M. Edwards, C. J. Williams, C. W. Clark, M. J.
Holland, P. A. Ruprecht and K. A. Burnett, Phys. Rev. A {\bf 54}, 661 (1996).
\bibitem{Ilinski} K. N. Ilinski and A. S. Stepanenko, Hydrodynamics of a Bose
condensate: beyond the mean field approximation, preprint (cond-mat/ 9607202)
1997.



\end{references}
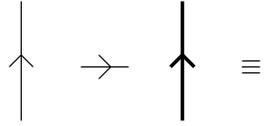
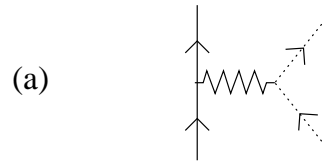
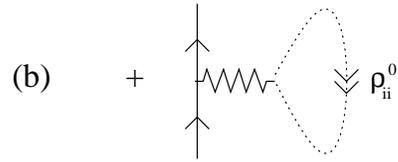
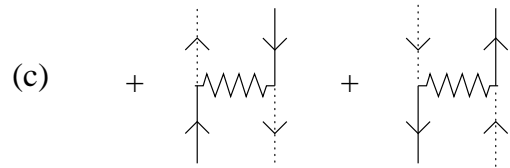
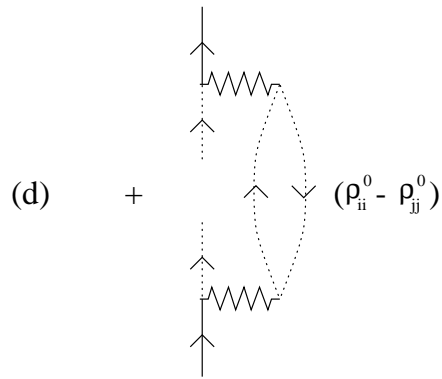
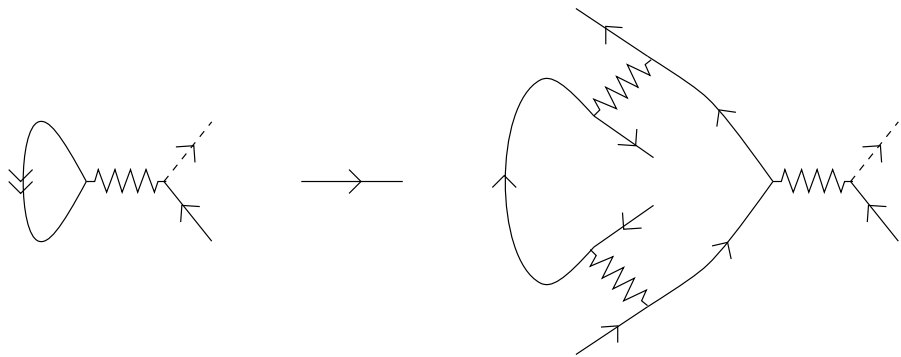
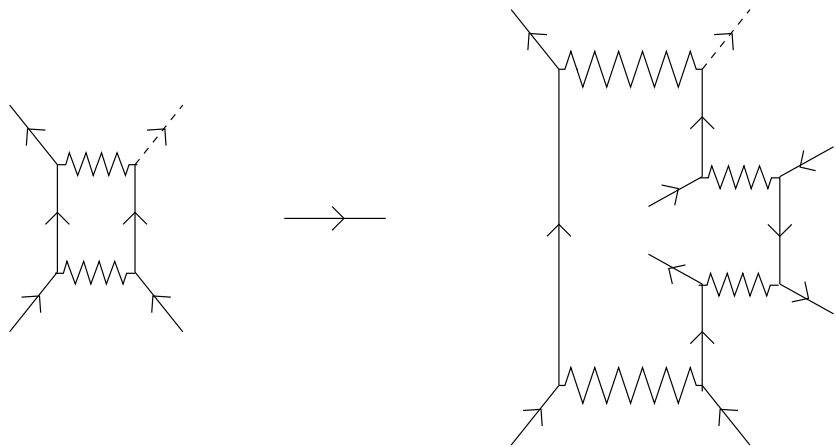
\end{document}